\documentclass[runningheads]{llncs}

\usepackage{xcolor}
\usepackage{enumitem}
\usepackage{verbatim}
\usepackage{colortbl} 

\usepackage[T1]{fontenc}
\usepackage{graphicx}
\usepackage{fancyhdr}
\fancypagestyle{firstpage}{
  \fancyhf{}
  
  \fancyhead[C]{\footnotesize Published in: Proceedings of the 28th Workshop on Requirements Engineering (WER2025). DOI: 10.29327/1588952.28-10}
}

\begin{document}

\title{Exploring LLMs for User Story Extraction from Mockups}
%
%

\author{Firmenich Diego\inst{1}*\and
Antonelli Leandro\inst{2,3} \and
Pazos Bruno\inst{1,4}\and
Lozada Fabricio\inst{3,6}\and
Morales Leonardo\inst{1,4,5}}

\authorrunning{Firmenich et al.}

%

\institute{Departamento de Informática, Facultad de Ingeniería, Universidad Nacional de la Patagonia, Argentina \and
LIFIA, Facultad de Informática, Universidad Nacional de La Plata, Argentina\and
CAETI, Facultad de Tecnología Informática - Universidad Abierta Interamericana\and
Laboratorio de Ciencias de la Imágenes,  DIEC, UNS\and
Instituto Patagónico de Ciencias Sociales y Humanas, CONICET-CENPAT\and
UNIANDES, Universidad Regional Autónoma de los Andes, Ecuador\\
\email{*dafirmenich@ing.unp.edu.ar}}

\maketitle              
\thispagestyle{firstpage}
\pagestyle{headings}

\begin{abstract}
User stories are one of the most widely used artifacts in the software industry to define functional requirements. In parallel, the use of high-fidelity mockups facilitate end-user participation in defining their needs. In this work, we explore how combining these techniques with large language models (LLMs) enables agile and automated generation of user stories from mockups. To this end, we present a case study that analyzes the ability of LLMs to extract user stories from high-fidelity mockups, both with and without the inclusion of a glossary of the Language Extended Lexicon (LEL) in the prompts. Our results demonstrate that incorporating the LEL significantly enhances the accuracy and suitability of the generated user stories. This approach represents a step forward in the integration of AI into requirements engineering, with the potential to improve communication between users and developers.

\end{abstract}

\section{Introduction}

In the dynamic landscape of software development, defining and communicating requirements is paramount to the success of any project \cite{hussain2016role}. Among the various techniques and artifacts employed for this purpose, user stories have emerged as one of the most widespread. Serving as concise, user-centric descriptions of functionality, user stories facilitate communication between technical teams and stakeholders, fostering collaboration and ensuring that software solutions are truly aligned with user needs and business goals \cite{rehkopf2021user}. Their widespread adoption is due to their ability to promote iterative development, facilitate clear communication, and ultimately drive the creation of valuable and user-focused software products throughout the industry \cite{lucassen2016use}. They are also technically simple artifacts containing text, minimally structured, and easy to manage with various tools. Other ways for end-users to specify requirements include mockups. Through mockups, users can specify their requirements with much greater precision in terms of UI \cite{filipovic2021rapid}. These techniques, used in web application development, are even more useful and intuitive when, through augmentation techniques, existing web applications are used as a canvas for high-fidelity mockup creation. However, ensuring consistency and correspondence between user stories and mockups is also a challenge \cite{kolthoff2024interlinking}. 

In this way, mockups, as artifacts, complement the definition of user stories, and user stories textually complement mockups. However, they are often insufficient on their own for comprehensive requirements management and developer implementation. This inherent limitation necessitates complementing visual mockups with textual or structured information, even in user-centered processes where users might provide this input directly through annotations or descriptions. Bridging this gap between visual specification and structured textual requirements is a recognized need in the field, explored by various approaches focused on extracting or complementing visual artifacts with textual requirements. \cite{ravid2000method,urbieta2018improving}\\

\indent In this work, we show how, based on recent advances in large language models (LLMs), it is possible to derive user stories from images corresponding to mockups. This has great value, as it allows for even more streamlined requirement definition with different artifacts. However, in the same way that traditionally not all users of a particular application specify their user stories with a common language. And this is a source of ambiguity \cite{amna2022ambiguity}. In this work, we test how, in addition to facilitating the derivation of text for user stories from images of their mockups, LLMs are capable of generating very precise user stories if a specific glossary is provided to them at the time of processing. Particularly, this proposal uses the Language Extended Lexicon (LEL) \cite{DBLP:conf/re/LeiteF93}. This adds another layer, not only facilitating the automatic generation of user stories from mockups, but also facilitating and guaranteeing the definition of these artifacts with the terminology that truly corresponds to each domain, since the LEL is originally built by requirements engineers in each domain, and consists of a series of very well-structured textual descriptions of the application domain.\\
\indent This work is organized as follows: Section 2 establishes the background, providing a synthesized review of the pillars that underpin the work: augmentation-based mockups, LLMs, and LEL. Likewise, a brief review of recent related works is made. Section 3 describes the proposed process in more detail. Section 4 shows the evaluation based on a case study, and Section 5 shows the results and discussion. Finally, Section 6 presents our conclusions and proposes future work.

\section{Background and related works}

High-fidelity mockups are interactive and detailed representations of user interfaces that play a crucial role in software development by enhancing communication among stakeholders and optimizing design validation \cite{Firmenich_Morales_Mura_Calfuquir_2024,samir2024model}.
These mockups, which closely resemble the final product, enable more effective user feedback and ensure that the final product aligns with expected requirements.
Advanced tools such as Mockplug \cite{Firmenich_Morales_Mura_Calfuquir_2024} and machine learning-based techniques, such as those used in SketchingInterfaces \cite{wimmer2020sketchinginterfaces}, have demonstrated improvements in development efficiency: Mockplug allows for a rapid technical requirement specification directly within the user interface, while SketchingInterfaces enables the automatic conversion of hand-drawn sketches into high-fidelity mockups.
Furthermore, automating code generation from these mockups significantly reduces both development costs and time, allowing developers to focus on functionality rather than repetitive design tasks \cite{samir2024model}.

Mockups complement textual requirement artifacts, such as user stories, by offering a visual representation of the intended functionality. However, maintaining consistency between user stories and mockups remains a challenge \cite{kolthoff2024interlinking}. Recent studies have explored the automation of traceability between user stories and graphical artifacts using LLMs \cite{kolthoff2024interlinking}. Our work builds on this synergy by exploring how LLMs can derive user stories from high-fidelity mockups while ensuring domain-specific accuracy through the integration of an glossary of the domain: the Language Extended Lexicon (LEL).\\
\indent The LEL provides a structured representation of linguistic symbols within a specific domain \cite{leite1993strategy}.
In Software Engineering, the LEL can be described as an enriched glossary that not only defines terms but also incorporates functions and behaviors, serving as a domain metamodel \cite{wehbe2011language}.
Its primary foundation lies in the premise that user and client engagement is strengthened when they share a common language with software engineers, thereby facilitating communication between stakeholders and developers.
Its application helps reduce ambiguity, improve alignment between initial requirements and user needs, and expand domain knowledge \cite{antonelli2024knowledge}. 
Although adequate use of the LEL glossary implies additional work, this extra activity can result in benefit in complex domains or teams without experienced workers, since the LEL glossary optimizes the elicitation of functional requirements from artifacts such as use cases and user stories, providing a more structured and precise representation of system expectations \cite{antonelli2012deriving}.
Moreover, it plays a key role in identifying and addressing non-functional requirements by treating them as first-class elements within the specification process, allowing for a more comprehensive approach to stakeholder demands \cite{neto2000non}.\\
\indent By integrating the LEL with LLMs, we leverage this structured linguistic knowledge to enhance the consistency and precision of automatically generated user stories.

\indent LLMs are Artificial Intelligence systems trained on enormous datasets and given their flexibility, with a very large amount of trainable parameters. 
Thanks to their architecture and multiple features, these models are capable of processing and responding to human-like text inputs by generating outputs with similar characteristics, appearing to understand the meaning of words and sentences given the context in which they are presented.  \\
A big part of these model’s success can be attributed to their architecture known as transformer \cite{Vaswani2017} which basically allow them to “pay attention” to specific portions in the input text and analyze the relationship with the rest of the input data; how much these portions are influenced by other data in the input and conversely how they affect the rest of the data. \\
With the goal of capturing as much information as possible, LLMs have millions or even billions of trainable parameters. 
During a pre-training stage \cite{devlin2018bert}, these models are trained on huge volumes of unlabeled data in text format \cite{Raffel2019,Zhu2015,li2023blip} in order to create and adjust their knowledge base providing the model with basic notions of the language, its structure, how it is composed and how context is built in it. In further stages, this knowledge base is extended through a process known as fine-tuning, allowing the model to be trained for specific purposes.
Nowadays, it is common to find LLMs behind many online services, such as: virtual assistants \cite{apple_siri,amazon_alexa}, chatbots \cite{openai_chatgpt,google_gemini}, automatic translation \cite{deepl_translator,google_translate}, content generation \cite{geniusee_generative_ai,onilab_llm_training}, and sentiment analysis among others.

LLMs have been increasingly explored in software engineering applications, including requirements engineering. Recent studies have investigated their ability to generate user stories, extract requirements from textual descriptions, and even refine ambiguous requirements \cite{amna2022ambiguity,jin2024llms}. For example, Kolthoff et al. \cite{kolthoff2024interlinking} proposed an LLM-based approach to interlink user stories with graphical user interface prototypes, demonstrating improvements in requirement traceability. Other works have explored the use of LLMs for requirements classification, ambiguity resolution \cite{belzner2023large}, and structured requirements generation \cite{wei2024requirements}.

Despite these advancements, relatively few studies have explored the direct derivation of user stories from high-fidelity mockups. Works such as Firmenich et al. \cite{firmenich2014platform} and Wimmer et al. \cite{wimmer2020sketchinginterfaces} have examined how augmented user interactions can facilitate requirement elicitation. Our work aims to address this gap by evaluating how LLMs can process visual representations of requirements and generate structured user stories, particularly when enhanced with an LEL. This approach builds upon prior work in automated requirements engineering and demonstrates how domain-specific lexicons can refine the output of LLMs.

\section{Contribution}
\begin{figure}
    \centering
    \includegraphics[width=0.8\linewidth]{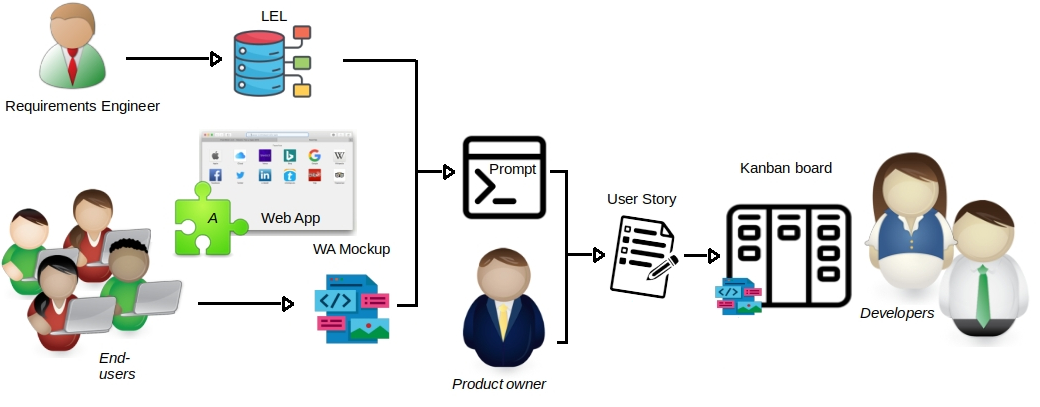}
    \caption{Proposed Collaborative Workflow}
    \label{fig:scenario}
\end{figure}

The proposed approach is illustrated in Fig. \ref{fig:scenario}, where the collaborative flow between the various actors involved in the proposal can be appreciated. End-users interact with a tool for high-fidelity mockup creation (identified in the figure as a browser add-on, that has a puzzle piece “A”). This allows them to define their requirements through three key actions: selecting relevant pre-existing elements for their needs, adding new components (widgets, elements from other parts of the application or other applications), and removing unwanted elements.

Using the web application as a canvas, the modified elements (selected or added) adopt a distinctive visual style, employing a characteristic hand-drawn font typical of mockups. This facilitates their identification compared to the elements of the original application.

In parallel, requirements engineers describe the language of the application domain, identifying and describing LEL symbols. The output produced by the requirements engineers is a structured glossary that textually describes all actors (subjects) in the application domain, with all their actions (verbs) and all concepts (objects) involved, along with their situations (states), establishing meaning (notion) for each of them as well as the relationships (behavioral responses) between them.\\
\indent All this valuable information for each requirement, that is, the screenshot of the original application, the mockup, and the LEL glossary, is used to feed the context of the LLM. Although the LEL glossary is not intuitive and needs some experience, it is not hard to learn and provides the knowledge necessary to complement the visual artifacts. The product owner, using a prompt along with this information, can then complement the definition of the original requirement, obtaining a user story appropriate for the mockup created. Finally, after managing and prioritizing the requirements, the product owner can manage these products, for example, in a Kanban board shared with the developers.

Table \ref{tab:table_roles} summarizes the roles of the proposed process, the actions that each actor must perform, as well as the tools involved, and the result and format obtained.

\begin{table}[h]
\centering
\caption{Roles, Actions, Tools, Products, and Formats in the Proposed Approach}
\renewcommand{\arraystretch}{1.5} 
\begin{tabular}{|p{2cm}|p{3cm}|p{2.5cm}|p{2cm}|p{2cm}|} 
\hline
\textbf{Role} & \textbf{Action} & \textbf{Tool} & \textbf{Product} & \textbf{Format} \\
\hline
End-user & Defines requirements & High-fidelity mockup tool & WA based Mockup & WA Mockup, Images \\
\hline
Requirements Engineer & Defines the application domain language & LEL edition Tool & LEL & Text \\
\hline
Product Owner & Manages and prioritizes requirements & Kanban, Prompt & Mockup+US & Text, Image \\
\hline
Developer & Implements requirements & Kanban, IDEs & Artifact & Software \\
\hline
\end{tabular}
\label{tab:table_roles}
\end{table}

To further streamline the process, once the end-user defines their requirements using the mockup tool and the mockups are finalized, the tool can automatically interact with the LLM via API. This interaction generates the corresponding user story, which can then be seamlessly integrated into requirement management platforms, such as a Kanban board managed by the product owner\cite{marticorena2023development}.

\section{Proof of Concept and Preliminary Validation}

This section presents two use cases to illustrate the proposal. The first one is based on a widely known application, while the second one focuses on a niche domain. 

For the first case, we selected YouTube, a widely used website where specifying a LEL was not considered necessary. This example showcases the power of LLMs in interpreting mockups independently, relying solely on the mockup itself without additional context.

\begin{figure}
    \centering
    \includegraphics[width=0.8\linewidth]{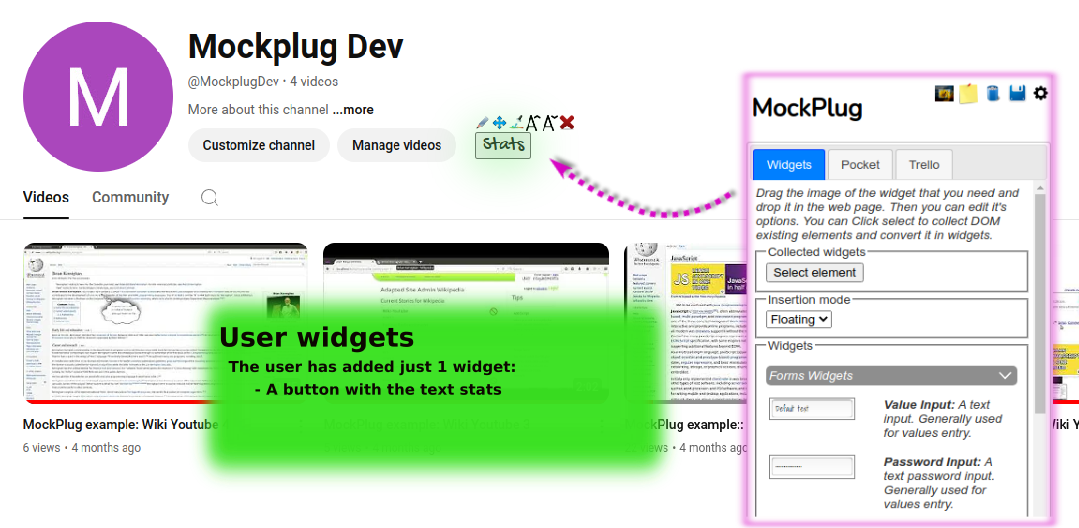}
    \caption{User Creating a High-Fidelity Mockup}
    \label{fig:user-mockup}
\end{figure}

In this straightforward example, the user has simply added a button to enable access to statistics from the channel view, as shown in Fig. \ref{fig:user-mockup}.

\indent It's important to note that this figure highlights the tool the user employed to add a button to an existing channel interface. The new button was added simply by dragging and dropping it onto the web application. Subsequently, the user only needed to edit its properties to set the text to “stats.”  Through this simple gesture, the user effectively communicated the desire for a new button, its intended location, and its desired label. The mockup visually distinguishes this new button by using a different font style. \\
\indent A screenshot of the high-fidelity mockup, along with a screenshot of the original version of the application before any modifications, was attached to the prompt. The prompt used is simply: \textit{'A YouTube user who posts videos, based on the original YouTube version (image 1), has made a high-fidelity mockup to describe a requirement (image 2). Could you make me a user story for that mockup?'}.

\begin{figure}
    \centering
    \includegraphics[width=0.8\linewidth]{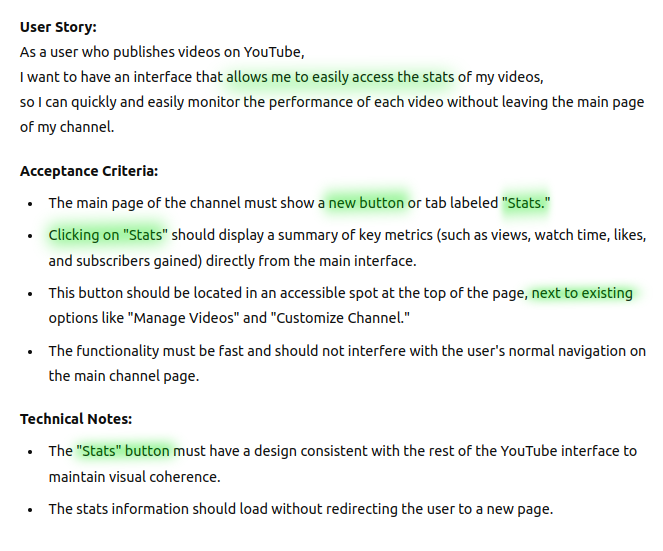}
    \caption{User Creating a High-Fidelity Mockup}
    \label{fig:user-story}
\end{figure}

In the Fig. \ref{fig:user-story}, it can be seen how the LLM provides a very suitable user story for the provided mockup. Although it is a very simple mockup, it can be appreciated how the LLM was able to interpret the image, distinguish that there was a new element, and that the element referred to statistics. The resulting user story explicitly states that a new button is desired, that it should be located next to the existing buttons, and that a consistent design is desired, making the new button look the same as the originals.

The second case study corresponds to an application with a very particular domain, such as the LeafLab application\cite{almonacid2019mobile}. LeafLab is a web application used on an intranet by biological scientists specializing in botany. We have used this application to represent case studies in previous works 
 \cite{marticorena2023development}.

\begin{figure}[h]
    \centering
    \includegraphics[width=0.8\linewidth]{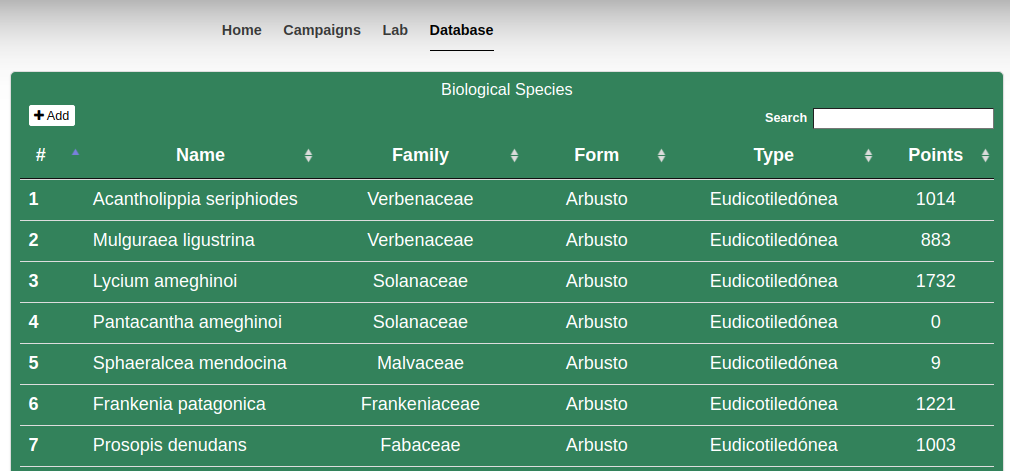}
    \caption{LeafLab web application displaying floristic species list, ordered by system \#id.}
    \label{fig:leaf-lab-original}
\end{figure}

In this example, we see how the biologist navigates the list of floristic species, as shown in Fig. \ref{fig:leaf-lab-original}. In this screenshot, it can be easily seen that the application offers a list of species, ordering them by the \textit{\#id} number in the application, and this piece of information is not relevant to the biologist.

The biologist wants the species to be listed sorted by relevance in the field, that is, showing on top of the list those species that have been observed most frequently, thus generating, by default, a ranking of predominance.

\begin{figure}[h]
    \centering
    \includegraphics[width=0.8\linewidth]{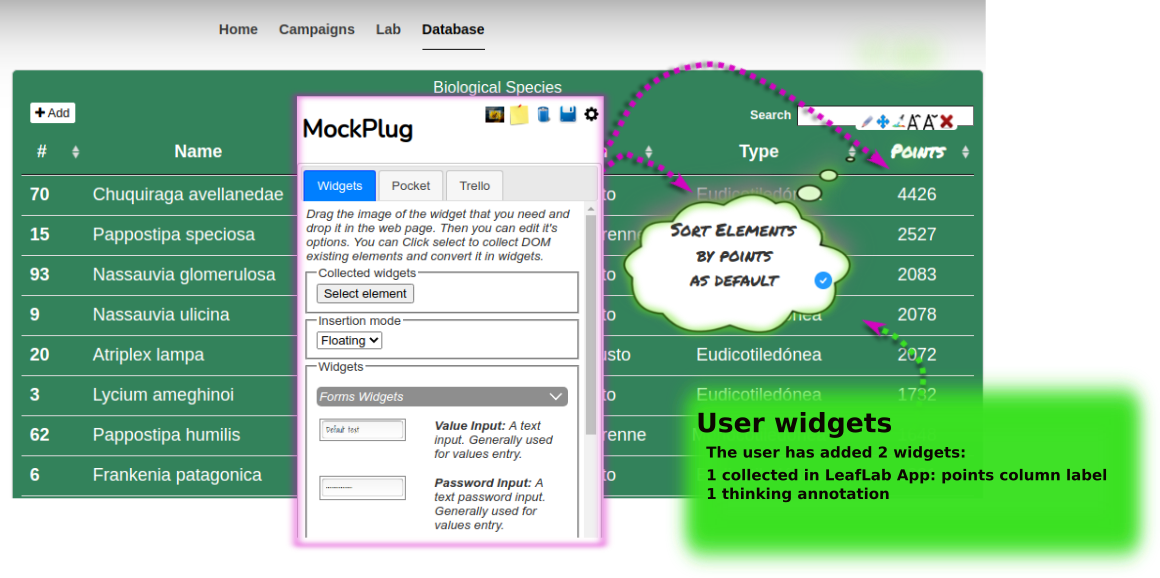}
    \caption{Biologist's mockup creation process for new LeafLab feature.}
    \label{fig:5-leaf-lab-original}
\end{figure}

To specify his need, he created a high-fidelity mockup as shown in Fig. \ref{fig:5-leaf-lab-original}. In his mockup, the species are listed by the number of points in which each species appeared in the exploration surveys.

\begin{figure}
    \centering
    \includegraphics[width=1\linewidth]{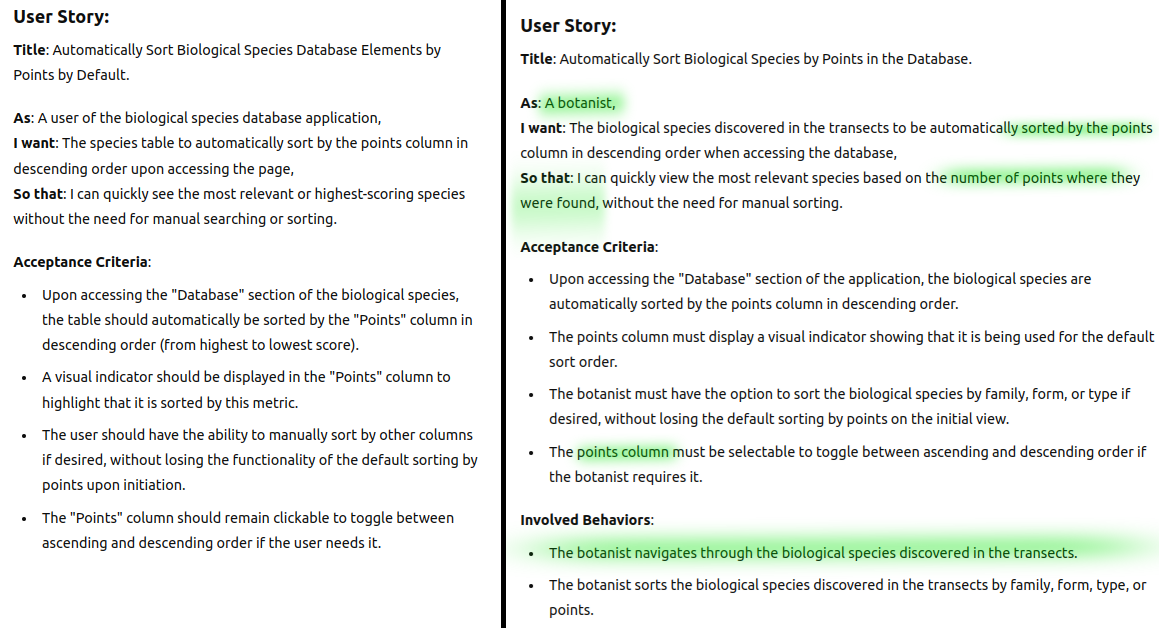}
    \caption{User story for the biologist mockpup, comparison without LEL (left) and with LEL (right) in the prompt.}
    \label{fig:67-leaflab-with-and-without-lel}
\end{figure}

It is important to note that the word \textit{points} is not just a simple score. For the biologist and within their specific domain, it has a very specific meaning: the listed species are those corresponding to thousands of specimens methodologically surveyed in the field during field campaigns. The word \textit{points} refer to the fact that in each survey, the biologist traverses an imaginary straight line of 300 meters, along which are distributed 100 equidistant sampling points. At each of these points in the field, the biologist takes a sample of a floristic specimen, which must be identified to species, often collecting a sample and using a stereoscopic microscope in the laboratory to differentiate one from another.\\
\indent To achieve this, taking the web application as a canvas, the biologist selects the existing title of the \textit{points} column using the mockup tool, converting it into an element of their mockup. Upon performing this action, that element is automatically displayed in a distinctive way from the others in the mockup, through a handwritten-style font. As can be seen in the figure, the user also adds an annotation to the mockup, in the form of a thought bubble, referencing that column, where they textually express \textit{Sort element by points as default}.\\
\indent In Fig. \ref{fig:67-leaflab-with-and-without-lel} on the left, we can observe how, when requesting the LLM to generate a user story based on the original version and the mockup, the user story, while surprisingly good, does not make precise use of the domain language. Specifically, when stating that the species should be sorted by the data in the \textit{points} column, it does so in terms of 'score' and not 'relevance', as in the explanatory sentence, it indicates: \textit{So that I can quickly see the most relevant or highest-scoring species without the need for manual searching or sorting}. While on the right, it illustrates how providing the LEL to the prompt enables the LLM to generate a user story that better aligns with the domain. The term 'points' no longer refers to a score but rather to the number of times a species was recorded in the surveyed field. The explanatory sentence is notably more precise and employs terminology specific to the application domain: \textit{So that I can quickly view the most relevant species based on the number of points where they were found, without the need for manual sorting.} \\

Notably, the distinctive elements introduced by the LEL are highlighted in the figure. It is also worth mentioning that the provided lexicon was incomplete, and only sufficient for this specific use case. As shown in Table \ref{tab:table_lel_subjects} and Table \ref{tab:table_lel_objects}. 

\begin{table}[h]
\centering
\caption{LEL portion for Subject, Notion, Behavioral responses}
\renewcommand{\arraystretch}{1.5} 
\begin{tabular}{|p{1.5cm}|p{2.5cm}|p{7.5cm}|} 
\hline
\textbf{Subject} & \textbf{Notion} & \textbf{Behavioral responses} \\
\hline
Biologist & User of the application licensed or doctorate in biological sciences and specialized in botany. & The biologist orders the species discovered in the transects by family. The biologist orders the species discovered in the transects by form. The biologist orders the species discovered in the transects by type. The biologist orders the species discovered in the transects by points. The biologist edits the properties of a species.\\
\hline
\end{tabular}
\label{tab:table_lel_subjects}
\end{table}

\begin{table}[h]
\centering
\caption{LEL portion for Objects and Notions}
\renewcommand{\arraystretch}{1.5} 
\begin{tabular}{|p{1.5cm}|p{10cm}|} 
\hline
\textbf{Object} & \textbf{Notion}\\
\hline
Transect & Imaginary straight line of approximately 300 meters, on which biologists sample specimens of biological species. Each transect is composed of 300 points. \\
\hline
Point & At each point of the 300 points of a transect, the biologist takes a sample of a specimen of a biological species found. \\
\hline
Biological Species & Species found in the field campaigns carried out by biologists. Each species may have been found in 0 or more points. \\
\hline
Visit & A traverse of a particular transect. Each traverse of a transect involves walking approximately 300 meters, recording the species of a particular floristic specimen every 3 meters, resulting in 100 points per transect. \\
\hline
\end{tabular}
\label{tab:table_lel_objects}
\end{table}

To preliminarily validate the proposed approach, we focus on the fundamental types of actions that a user can perform when constructing their high-fidelity mockup on the existing web application. That is, when building their mockup within the proposed approach, the user will specify their requirement based on some combination of the following possibilities:

\begin{itemize}[leftmargin=1.5cm]
    \item A. Add a new element (not previously existing).
    \item B. Add an existing element to the requirement.
    \item C. Remove existing elements.
    \item D. Compose elements from other parts of the application.
    \item E. Compose elements from other applications.
\end{itemize}

For each of these options, we conducted tests with and without the LEL, evaluating whether the LLM is capable of generating an appropriate user story for the mockup and with what level of precision it does so. Following the methodology used in the previous examples, a mockup was designed that incorporated different categories of actions, and the adequacy of the generated user story was evaluated. It should be mentioned that both the LEL and the mockups used in the validation were developed by the team responsible for the application, with experience in both the domain and the technologies involved.

Each user story was assigned a score according to the following value scale:
\begin{enumerate}[leftmargin=1.5cm]
    \item Does not distinguish any element of the requirement.
    \item Distinguishes only partially the elements that compose the requirement.
    \item Distinguishes the elements of the requirement, but does not describe them adequately.
    \item Distinguishes the elements of the requirement and describes some of them adequately.
    \item Distinguishes the elements of the requirement and describes them adequately in their entirety.
\end{enumerate}

The table summarizes the results, where the score with LEL is greater than or equal to the score without LEL, indicating the preliminary importance of using the glossary LEL. 

\begin{table}[h]
\centering
\caption{Type of Requirement Scores With and Without LEL}
\begin{tabular}{|c|c|c|}
\hline
\textbf{Type of Requirement} & \textbf{Score without LEL} & \textbf{Score with LEL} \\
\hline
A & 3 & 5 \\
\hline
B & 3 & 5 \\
\hline
C & 4 & 4 \\
\hline
D & 3 & 5 \\
\hline
E & 3 & 5 \\
\hline
\end{tabular}
\label{tab:requirement_scores}
\end{table}

\section{Conclusion}
\indent The rapid advancement of language models in recent years has continuously redefined multiple processes across diverse application domains, and this trend is expected to continue. The capabilities of these new tools continue to expand, driven by rapid advances in the generative AI industry.

Although numerous powerful tools have emerged to assist software engineers, the core processes of software engineering continue to evolve alongside these advancements.

In this work, we propose the integration of high-fidelity mockups and the glossary LEL to generate user stories for the definition of requirements using an LLM, and we provide evidence supporting its feasibility. Our results are highly encouraging: deriving user stories from mockups and leveraging a glossary LEL enable agile and precise involvement of end-users in requirement definition, while streamlining communication through existing web applications and appropriate tools. By incorporating the glossary LEL into the prompt before generating user stories, we enhance the value of domain analysis performed by requirements engineers, ensuring terminology consistency, and improving requirement clarity.

This approach has the potential to significantly impact software development workflows by reducing ambiguity in requirement definition and speeding up the initial phases of development. However, further validation is needed to assess its adaptability to diverse application domains and its robustness in handling complex requirements. Addressing potential biases in language models and refining the automation process to ensure alignment with stakeholder expectations remain key challenges.

Future work will focus on refining this integration, expanding its applicability, and conducting empirical studies to measure its effectiveness in real-world software development scenarios. Furthermore, exploring hybrid approaches that combine AI-generated content with human validation could further enhance the reliability of this method.

%
%
%


%
%
%
%

\bibliographystyle{splncs04}
\bibliography{bibliography}

@inproceedings{DBLP:conf/re/LeiteF93,
  author       = {Julio C{\'{e}}sar Sampaio do Prado Leite and
                  Ana Paula M. Franco},
  title        = {A strategy for conceptual model acquisition},
  booktitle    = {Proceedings of {IEEE} International Symposium on Requirements Engineering,
                  {RE} 1993, San Diego, California, USA, January 4-6, 1993},
  pages        = {243--246},
  publisher    = {{IEEE} Computer Society},
  year         = {1993},
  url          = {https://doi.org/10.1109/ISRE.1993.324851},
  doi          = {10.1109/ISRE.1993.324851},
  bibsource    = {dblp computer science bibliography, https://dblp.org}
}

@article{Vaswani2017,
   author = {Ashish Vaswani and Noam Shazeer and Niki Parmar and Jakob Uszkoreit and Llion Jones and Aidan N. Gomez and Łukasz Kaiser and Illia Polosukhin},
   isbn = {1706.03762v7},
   issn = {10495258},
   journal = {Advances in Neural Information Processing Systems},
   month = {6},
   pages = {5999-6009},
   publisher = {Neural information processing systems foundation},
   title = {Attention Is All You Need},
   volume = {2017-December},
   url = {https://arxiv.org/abs/1706.03762v7},
   year = {2017},
}

@article{Raffel2019,
   author = {Colin Raffel and Noam Shazeer and Adam Roberts and Katherine Lee and Sharan Narang and Michael Matena and Yanqi Zhou and Wei Li and Peter J. Liu},
   issn = {15337928},
   journal = {Journal of Machine Learning Research},
   month = {10},
   publisher = {Microtome Publishing},
   title = {Exploring the Limits of Transfer Learning with a Unified Text-to-Text Transformer},
   volume = {21},
   url = {https://arxiv.org/abs/1910.10683v4},
   year = {2019},
}

@article{Zhu2015,
   author = {Yukun Zhu and Ryan Kiros and Richard Zemel and Ruslan Salakhutdinov and Raquel Urtasun and Antonio Torralba and Sanja Fidler},
   month = {6},
   title = {Aligning Books and Movies: Towards Story-like Visual Explanations by Watching Movies and Reading Books},
   year = {2015},
}

@article{hussain2016role,
title={The role of requirements in the success or failure of software projects},
author={Hussain, Azham and Mkpojiogu, Emmanuel OC and Kamal, Fazillah Mohmad},
journal={International Review of Management and Marketing},
volume={6},
number={7},
pages={306--311},
year={2016},
publisher={{\.I}lhan {\"O}ZT{\"U}RK}
}

@article{rehkopf2021user,
title={User stories with examples and a template},
author={Rehkopf, Max},
journal={Atlassian. URL: https://www. atlassian. com/agile/project-management/user-stories [accessed 2021-12-03]},
year={2021}
}

@inproceedings{lucassen2016use,
title={The use and effectiveness of user stories in practice},
author={Lucassen, Garm and Dalpiaz, Fabiano and Werf, Jan Martijn EM van der and Brinkkemper, Sjaak},
booktitle={Requirements Engineering: Foundation for Software Quality: 22nd International Working Conference, REFSQ 2016, Gothenburg, Sweden, March 14-17, 2016, Proceedings 22},
pages={205--222},
year={2016},
organization={Springer}
}

@article{filipovic2021rapid,
title={Rapid Requirements Elicitation of Enterprise Applications Based on Executable Mockups},
author={Filipovi{\'c}, Milorad and Vukovi{\'c}, {\v{Z}}eljko and Dejanovi{\'c}, Igor and Milosavljevi{\'c}, Gordana},
journal={Applied Sciences},
volume={11},
number={16},
pages={7684},
year={2021},
publisher={MDPI}
}

@article{kolthoff2024interlinking,
title={Interlinking user stories and GUI prototyping: A semi-automatic LLM-based approach},
author={Kolthoff, Kristian and Kretzer, Felix and Bartelt, Christian and Maedche, Alexander and Ponzetto, Simone Paolo},
journal={arXiv preprint arXiv:2406.08120},
year={2024}
}

@article{marticorena2023development,
  title={Development iterations based on web augmentation and %context tasks},
  author={Marticorena, Lucy Gutierrez and Morales, Leonardo A and Antonelli, Leandro and Rossi, Gustavo and Firmenich, Diego},
  journal={Multimedia Tools and Applications},
 volume={82},
  number={8},
  pages={11793--11817},
  year={2023},
  publisher={Springer}
}

@inproceedings{almonacid2019mobile,
  title={Mobile and wearable computing in patagonian wilderness},
  author={Almonacid, Samuel and Klagges, Mar{\'\i}a R and Navarro, Pablo and Morales, Leonardo and Pazos, Bruno and Puigb{\'o}, Alexandra Contreras and Firmenich, Diego},
  booktitle={Cloud Computing and Big Data: 7th Conference, JCC\&BD 2019, La Plata, Buenos Aires, Argentina, June 24--28, 2019, Revised Selected Papers 7},
  pages={137--154},
  year={2019},
  organization={Springer}
}

@article{amna2022ambiguity,
  title={Ambiguity in user stories: A systematic literature review},
  author={Amna, Anis R and Poels, Geert},
  journal={Information and Software Technology},
  volume={145},
  pages={106824},
  year={2022},
  publisher={Elsevier}
}

@article{Firmenich_Morales_Mura_Calfuquir_2024, title={Mockplug: A high-fidelity mocking tool for plugging functional requirements into existing web applications}, volume={7}, DOI={10.24294/csma.v7i1.6736}, number={1}, journal={Computer software and media applications}, publisher={EnPress Publisher}, author={Firmenich, Diego Andrés and Morales, Leonardo and Mura, Gastón and Calfuquir, Nicolás}, year={2024}, month=sep, pages={6736} }

@article{samir2024model,
title={A Model for Automatic Code Generation from High Fidelity Graphical User Interface Mockups using Deep Learning Techniques.},
author={Samir, Michel and Elsayed, Ahmed and Marie, Mohamed I},
journal={International Journal of Advanced Computer Science \& Applications},
volume={15},
number={3},
year={2024}
}

@inproceedings{wimmer2020sketchinginterfaces,
title={SketchingInterfaces: a tool for automatically generating high-fidelity user interface mockups from hand-drawn sketches},
author={Wimmer, Christoph and Untertrifaller, Alex and Grechenig, Thomas},
booktitle={Proceedings of the 32nd Australian Conference on Human-Computer Interaction},
pages={538--545},
year={2020}
}

@inproceedings{leite1993strategy,
title={A strategy for conceptual model acquisition},
author={Leite, JCS do P and Franco, Ana Paula M},
booktitle={[1993] Proceedings of the IEEE International Symposium on Requirements Engineering},
pages={243--246},
year={1993},
organization={IEEE}
}

@inproceedings{wehbe2011language,
title={The Language Extended Lexicon, Revisited},
author={Wehbe, Ricardo},
booktitle={XII Argentine Symposium on Software Engineering (ASSE 2011)(XL JAIIO, C{\'o}rdoba, 1{\textordmasculine} y 2 de septiembre de 2011)},
year={2011}
}

@article{antonelli2024knowledge,
title={Knowledge Extraction from the Language Extended Lexicon Glossary Using Natural Language Processing},
author={Antonelli, Leandro and Lezoche, Mario and Delle Ville, Juliana},
journal={TecnoL{\'o}gicas},
number={59},
pages={2},
year={2024},
publisher={Instituto Tecnol{\'o}gico Metropolitano}
}

@inproceedings{antonelli2012deriving,
title={Deriving requirements specifications from the application domain language captured by Language Extended Lexicon.},
author={Antonelli, Leandro and Rossi, Gustavo and do Prado Leite, Julio Cesar Sampaio and Oliveros, Alejandro},
booktitle={WER},
year={2012}
}

@inproceedings{neto2000non,
title={Non-Functional Requirements for Object-Oriented Modeling.},
author={Neto, Jaime De Melo Sabat and do Prado Leite, Julio Cesar Sampaio and Cysneiros Filho, Luiz M{\'a}rcio},
booktitle={WER},
pages={109--125},
year={2000}
}

@article{devlin2018bert,
  title={Bert: Pre-training of deep bidirectional transformers for language understanding},
  author={Devlin, Jacob},
  journal={arXiv preprint arXiv:1810.04805},
  year={2018}
}

@inproceedings{li2023blip,
  title={Blip-2: Bootstrapping language-image pre-training with frozen image encoders and large language models},
  author={Li, Junnan and Li, Dongxu and Savarese, Silvio and Hoi, Steven},
  booktitle={International conference on machine learning},
  pages={19730--19742},
  year={2023},
  organization={PMLR}
}

@misc{apple_siri,
  author = {{Apple Inc.}},
  title = {{Siri}},
  howpublished = {\url{https://www.apple.com/siri/}},
  note = {Last access: 2025-02-17}
}

@misc{amazon_alexa,
  author = {{Amazon.com Inc.}},
  title = {{Amazon Alexa}},
  howpublished = {\url{https://www.alexa.com/}},
  note = {Last access: 2025-02-17}
}

@misc{openai_chatgpt,
  author = {{OpenAI}},
  title = {{ChatGPT}},
  howpublished = {\url{https://openai.com/index/chatgpt/}},
  note = {Last access: 2025-02-17}
}

@misc{google_gemini,
  author = {{Google LLC.}},
  title = {{Gemini}},
  howpublished = {\url{https://gemini.google.com/}},
  note = {Last access: 2025-02-17}
}

@misc{deepl_translator,
  author = {{DeepL GmbH.}},
  title = {{DeepL Translator}},
  howpublished = {\url{https://www.deepl.com/en/translator}},
  note = {Last access: 2025-02-17}
}

@misc{google_translate,
  author = {{Google LLC.}},
  title = {{Google Translate}},
  howpublished = {\url{https://translate.google.com}},
  note = {Last access: 2025-02-17}
}

@misc{geniusee_generative_ai,
  author = {{Geniusee Inc.}},
  title = {{Generative AI development services}},
  howpublished = {\url{https://geniusee.com/generative-ai-development}},
  note = {Last access: 2025-02-17}
}

@misc{onilab_llm_training,
  author = {{Onilab Inc.}},
  title = {{LLM training and development}},
  howpublished = {\url{https://onilab.com/services/llm-training-and-development}},
  note = {Last access: 2025-02-17}
}

@inproceedings{firmenich2014platform,
  title={A platform for web augmentation requirements specification},
  author={Firmenich, Diego and Firmenich, Sergio and Rivero, Jos{\'e} Mat{\'\i}as and Antonelli, Leandro},
  booktitle={Web Engineering: 14th International Conference, ICWE 2014, Toulouse, France, July 1-4, 2014. Proceedings 14},
  pages={1--20},
  year={2014},
  organization={Springer}
}

@article{jin2024llms,
  title={From llms to llm-based agents for software engineering: A survey of current, challenges and future},
  author={Jin, Haolin and Huang, Linghan and Cai, Haipeng and Yan, Jun and Li, Bo and Chen, Huaming},
  journal={arXiv preprint arXiv:2408.02479},
  year={2024}
}

@article{wei2024requirements,
  title={Requirements are All You Need: From Requirements to Code with LLMs},
  author={Wei, Bingyang},
  journal={arXiv preprint arXiv:2406.10101},
  year={2024}
}

@inproceedings{belzner2023large,
  title={Large language model assisted software engineering: prospects, challenges, and a case study},
  author={Belzner, Lenz and Gabor, Thomas and Wirsing, Martin},
  booktitle={International Conference on Bridging the Gap between AI and Reality},
  pages={355--374},
  year={2023},
  organization={Springer}
}

@inproceedings{urbieta2018improving,
  title={Improving mockup-based requirement specification with end-user annotations},
  author={Urbieta, Matias and Torres, Nahime and Rivero, Jos{\'e} Matias and Rossi, Gustavo and Dominguez-Mayo, Francisco Jost{\'o}},
  booktitle={Agile Processes in Software Engineering and Extreme Programming: 19th International Conference, XP 2018, Porto, Portugal, May 21--25, 2018, Proceedings 19},
  pages={19--34},
  year={2018},
  organization={Springer International Publishing}
}

@article{ravid2000method,
  title={A method for extracting and stating software requirements that a user interface prototype contains},
  author={Ravid, Alon and Berry, Daniel M},
  journal={Requirements Engineering},
  volume={5},
  pages={225--241},
  year={2000},
  publisher={Springer}
}
\end{document}